\documentstyle[aps,epsf]{revtex}

\preprint{NUC-MINN-01/6-T}

\newcommand{\pslash}{\not{\! p}}

\newcommand{\be}{\begin{equation}}
\newcommand{\ee}{\end{equation}}
\newcommand{\ba}{\begin{eqnarray}}
\newcommand{\ea}{\end{eqnarray}}

\begin{document}
\draft

\title{Color superconductivity and nondecoupling
phenomena in (2+1)-dimensional QCD}

\author{V. A. Miransky\thanks{
        On leave of absence from
        Bogolyubov Institute for Theoretical Physics,
        Kiev 252143, Ukraine.}}

\address{Department of Applied Mathematics,
University of Western Ontario,
London, Ontario N6A 5B7, Canada}

\author{G.W. Semenoff}

\address{Department of Physics and Astronomy,
        University of British Columbia,
        6224 Agricultural Road, Vancouver,
        British Columbia V6T 1Z1, Canada.}

\author{I.A. Shovkovy\thanks{
        On leave of absence from
        Bogolyubov Institute for Theoretical Physics,
        Kiev 252143, Ukraine.}}

\address{School of Physics and Astronomy,
         University of Minnesota,
         Minneapolis, MN 55455}

\author{L.C.R. Wijewardhana}

\address{Physics Department,
         University of Cincinnati,
         Cincinnati, Ohio 45221-0011}

\date{\today}
\maketitle

\begin{abstract} 
The possibility of generating a color superconducting state in
$(2+1)$-dimensional QCD is analyzed. The gap equation in the leading,
hard-dense loop improved, one-gluon exchange approximation is derived and
solved. The magnitude of the order parameter is proportional to a power of
the coupling constant. For an asymptotically large chemical potential, a
qualitatively new [with respect to the $(3+1)$-dimensional case] 
phenomenon of nondecoupling of the fermion pairing dynamics from the
infrared one is revealed and discussed.  
\end{abstract}




\section{Introduction}
\label{introduction}

The study of color superconductivity has attracted a great deal of
attention in the past few years. The first theoretical studies of the
subject appeared over twenty years ago \cite{BarFra,Bail}. For many
reasons, however, its importance was not fully appreciated until recently
\cite{W1,S1}. The last couple of years were marked by intensive
theoretical studies, producing further understanding of the subtleties of
physics behind color superconductivity
\cite{PR1,Son,us,SW2,PR2,H1,Br1,CFL,cnt,us3,us2,Spt,Rsch,CaD,CGH,SSt,RW4,HZB,Zar,Be3,RSS,neu,Ngu}
(see Refs.~\cite{Rev-RW,Rev-A} for a review).

The color superconducting phase is expected to appear in sufficiently
dense quark matter. It is quite likely, therefore, that it exists inside
some neutron stars \cite{HIC-Stars}. Of course, the hope is that the
superconducting order parameter is large enough to result in clear
observational signatures. While this still remains to be confirmed, more
theoretical work is needed before any solid quantitative predictions would
be available.

In this paper, we study the color superconducting phase in
$(2+1)$-dimensional QCD at zero temperature and finite density. This
theory is superrenormalizable and, therefore, asymptotically free. The
conventional wisdom is that at asymptotically high quark density the
dynamics in asymptotically free theories is weakly interacting and
therefore drastically simplified. This is the case in $(3+1)$-dimensional
QCD \cite{Son,us,SW2,PR2}. One of our goals is to check the validity of
this picture in $(2+1)$ dimensions. We study the problem of color
superconductivity using the conventional method of the Schwinger-Dyson
(SD) equation in the hard-dense loop (HDL) improved one-gluon exchange
approximation. By solving the SD equation analytically, we obtain an
approximate solution for the color superconducting gap. The result is
proportional to a power of the coupling constant. This contrasts the
situation in $(3+1)$-dimensional QCD, where the expression for the gap is
nonanalytic in the coupling constant [$\Delta \sim \mu
\exp(-C/\sqrt{\alpha_{s}})$ with $\mu$ being the quark chemical potential
and $\alpha_s \equiv \alpha_{s}(\mu)$] when the coupling $\alpha_{s}$ is
weak \cite{PR1,Son,us,SW2,PR2}.

Is the conventional wisdom about weakly interacting dynamics at high quark
density in $(2+1)$-dimensional QCD justified? A detailed analysis of the
solution of the SD equation in this theory reveals a qualitatively new
[with respect to the $(3+1)$-dimensional case] phenomenon: nondecoupling
of the dynamics of the fermion pairing from the nonperturbative infrared
one, even in the case of asymptotically large values of the chemical
potential. We argue that this phenomenon is common in lower dimensional
models. The relevance of the $1/N_{f}$ expansion, with $N_{f}$ the number
of fermion flavors, for solving difficulties induced by this phenomenon is
pointed out.

This paper is organized as follows. In Sec.~\ref{masses}, we discuss
different types of the Dirac and Majorana masses, paying special attention
to their transformation properties under parity which itself could be
defined in several ways. We argue that, in the QCD model at hand, only
that color breaking Majorana mass is generated which preserves the global
$SU(2)$ ``pre-flavor" symmetry. The hard-dense loop gluon propagator is
derived in Sec.~\ref{HDL-sec}. Then, in Sec.~\ref{SD-sec}, we derive the
SD equation in the HDL improved rainbow approximation. We solve the
equation analytically and present an approximate solution. In
Sec.~\ref{nondecoupling} the nondecoupling phenomenon taking place for
large values of the chemical potential in $(2+1)$-dimensional QCD is
discussed. Our conclusions are given in Sec.~\ref{conclusion}. In
Appendix~\ref{notation}, we describe our notation and list some useful
formulas, used throughout the main text. And in Appendix~\ref{AppB},
we give an approximate analytical solution of the gap equation.

\section{Ordinary and Majorana mass terms}
\label{masses}

The $(2+1)$-dimensional models are somewhat different from their analogues
in $(3+1)$ dimensions. In particular, the global ``flavor" symmetries have
somewhat different status. From studying color superconductivity in
different models, we know that often color breaking is accompanied by
locking with flavor symmetries \cite{CFL,Spt}. Before studying the SD
equation for a possibility of  color superconducting gap, we discussed
some general symmetry properties of all possible Dirac and Majorana
masses in $(2+1)$-dimensional QCD.

In $(2+1)$ dimensions, the irreducible spinor representation is two
dimensional. For many purposes, however, it is more convenient to work
with the four component spinors that contain two different irreducible
representations. To make a connection with the usual terminology in
$(3+1)$ dimensions, we could assign a notion of flavor to each four
component spinor, while using ``pre-flavor" for two component irreducible
spinors.

In the case of four component spinors, the kinetic
term of quarks,
\be
{\cal L}_{kin}=\bar{\psi}\left(i\gamma^{\mu}D_{\mu}
+\mu \gamma^{0}\right)\psi,
\ee
has $U(1)_{V} \times SU(2)$ (pre-flavor) symmetry
\cite{JacTem,DJT,P84,ABKW}, generated by
\be
{\cal I},\quad
T^1=\gamma^{3} \gamma^{5} ,\quad
T^2=i\gamma^{3} , \quad
T^3=\gamma^{5}.
\ee
One could also add a Dirac mass term that respects this symmetry,
\be
m_1 \bar{\psi}T^{1}\psi.
\label{m1}
\ee
This mass, however, is odd under parity
\be
\psi(x_1, x_2) \to \gamma^{1} \psi(-x_1, x_2),
\label{parity}
\ee
and, furthermore, there are no modification of the parity transformation
(there are at least three other possible definitions,
$\psi\to\gamma^{0}\gamma^{2}\psi^{\prime}$,
$\psi\to\gamma^{5}\gamma^{1}\psi^{\prime}$ and
$\psi\to\gamma^{3}\gamma^{1}\psi^{\prime}$, respectively) that would make
the mass term in Eq.~(\ref{m1}) parity even. Also, having the mass term as
in Eq.~(\ref{m1}), would lead to the generation of the Chern-Simons term
at one loop order.

There are no other mass terms which would respect the $U(1)_{V} \times
SU(2)$ (pre-flavor) symmetry \cite{JacTem,DJT,P84,ABKW,ANW,AN}. For
example, each of the following mass terms
\be
m_0 \bar{\psi}\psi,\quad
i m_2 \bar{\psi}T^{2}\psi,\quad
i m_3 \bar{\psi}T^{3}\psi,
\label{ms}
\ee
breaks the symmetry down to $U(1)_{V}\times U(1)$, generated by $({\cal
I},T^{1})$, $({\cal I},T^{3})$, and $({\cal I},T^{2})$, respectively.
Note that the last two terms are even under parity defined in
Eq.~(\ref{parity}).  Although no definition of parity exists that would be
respected by all three mass terms in Eq.~(\ref{ms}), a different
definition of parity ($\psi\to\gamma^{5}\gamma^{1}\psi^{\prime}$ or
$\psi\to\gamma^{3}\gamma^{1}\psi^{\prime}$) could make the first term in
Eq.~(\ref{ms}) parity even. Also, since the $SU(2)$ transformations could
change each term in Eq.~(\ref{ms}) into another, all three of them are
physically equivalent.

As a result of spontaneous breaking of the color symmetry, different
kinds of Majorana masses could be generated. In analogy with the
ordinary masses, the following four structures are possible:
\begin{mathletters}
\ba
\Delta \bar{\psi}^{C}\psi, &\quad & U(1)
\quad \mbox{by} \quad T^{2},   \label{d0}\\
\Delta_{1} \bar{\psi}^{C}T^{1}\psi, &\quad & U(1)
\quad \mbox{by} \quad T^{3},   \label{d1}\\
\Delta_{2} \bar{\psi}^{C}T^{2}\psi, &\quad & SU(2)
\quad \mbox{by} \quad T^{1}, T^{2}, T^{3} \label{d2}\\
\Delta_{3} \bar{\psi}^{C}T^{3}\psi, &\quad & U(1)
\quad \mbox{by} \quad T^{1},  \label{d3}
\ea
\label{Maj}
\end{mathletters}
where $\psi^{C} \equiv C \bar{\psi}^{T}$ denotes the charge conjugate
spinor, and $C$ is a charge conjugation matrix, defined by $C^{-1}
\gamma_{\mu}C =-\gamma_{\mu}^{T}$ and $C=-C^{T}$. In Eqs.~(\ref{d0}) --
(\ref{d3}) we also indicated the symmetry of the Majorana mass terms and the
generators of the corresponding groups. Notice that the original
$U(1)_{V}$ is always broken.

The parity properties of the Dirac and Majorana mass terms are summarized
in Table~I.

\begin{center}
\centerline{TABLE I. Parity properties of Dirac and Majorana mass terms.}

\begin{tabular}{|c|c|c|c|c|}
\hline 
\hline
\hspace*{15mm} & \hspace*{15mm} & & \hspace*{15mm}& \\[-3mm]
 & $P_{1}=\gamma^{1}$  & $P_{2}=\gamma^{0}\gamma^{2}$
& $P_{3}=\gamma^{5}\gamma^{1}$  & $P_{4}=\gamma^{3}\gamma^{1}$ \\
\hline
 & & & & \\[-3mm]
$\bar{\psi}\psi$ & $-$ & $-$ & $+$ & $+$ \\
\hline
 & & & & \\[-3mm]
$\bar{\psi}T^{1}\psi$ & $-$ & $-$ & $-$ & $-$ \\
\hline
 & & & & \\[-3mm]
$\bar{\psi}T^{2}\psi$ & $+$ & $-$ & $+$ & $-$ \\
\hline
 & & & & \\[-3mm]
$\bar{\psi}T^{3}\psi$ & $+$ & $-$ & $-$ & $+$ \\
\hline
 & & & & \\[-3mm]
$\bar{\psi}^{C}\psi$ & $+$ & $-$ & $+$ & $+$ \\
\hline
 & & & & \\[-3mm]
$\bar{\psi}^{C}T^{1}\psi$ & $+$ & $-$ & $-$ & $-$ \\
\hline
 & & & & \\[-3mm]
$\bar{\psi}^{C}T^{2}\psi$ & $-$ & $-$ & $+$ & $-$ \\
\hline
 & & & & \\[-3mm]
$\bar{\psi}^{C}T^{3}\psi$ & $-$ & $-$ & $-$ & $+$ \\
\hline
\hline
\end{tabular}
\end{center}

While the original field $\psi$ belongs to the fundamental representation
of the (pre-flavor) $SU(2)$ group, the charge conjugate field $\psi^{C}$
belongs, in general, to the anti-fundamental representation. It is clear
than that the diquark fields such as the Majorana mass terms in
Eq.~(\ref{Maj}), could belong either to the (antisymmetric) singlet
representation ($\Delta_{2}$), or to the (symmetric) triplet one
($\Delta$, $\Delta_{1}$ and $\Delta_{3}$).

Since the most attractive diquark channel is antisymmetric (e.g.,
antitriplet in the case of $SU(3)$ color group), it becomes obvious that
the color symmetry breaking order parameter would be $SU(2)$ singlet as
that in Eq.~(\ref{d2}). Note that this order parameter is odd under the
parity defined in Eq.~(\ref{parity}), but even under the parity $P_{3}$
(see Table~I).

\section{Hard dense loop approximation for the gluon propagator}
\label{HDL-sec}

In dense media, the polarization effects play very important role. Below
we derive the polarization tensor in the so-called hard dense loop (HDL)
approximation. This approximation corresponds to dealing with semi-hard
(external) gluon lines, $\alpha \ll k \ll \mu$, and hard (internal)
fermion lines. Note that there are no one loop corrections with internal
gluon and ghost lines in HDL approximation. Our notation and some
formulas used in this section are given in Appendix~\ref{notation}.

The analytical expression for the polarization tensor reads
\be
\Pi ^{\mu \nu AB}(k)= 4 i \pi \alpha N_{f}
\int \frac{d q_0 d^2 q}{(2\pi )^{3}} \mbox{tr}
\left[ \gamma ^{\mu } T^{A} G(q) \gamma ^{\nu }
T^{B} G(q+k)\right] ,
\label{Pi-def}
\ee
where $\alpha$ is a (dimensionful) coupling constant of
$(2+1)$-dimensional QCD and $N_{f}$ counts the number of the
four-component quark flavors, not the pre-flavors.

In derivation of the HDL polarization tensor, it is sufficient to consider
only massless quarks. The corresponding propagator is given by
\be
G^{(0)}(q)=i\gamma^{0}\left[
\frac{1}{q_0+\epsilon^{+}_{q}-i\varepsilon } \Lambda^{(+)}_{q}
+\frac{1}{q_0-\epsilon^{-}_{q}
+i \varepsilon \mbox{ sign}(\epsilon^{-}_{q})}
\Lambda^{(-)}_{q} \right].
\ee
where $\epsilon_{p} ^{\pm }=|\vec{p}|\pm \mu $ and the projectors
$\Lambda^{(\pm )}_{p}$ are defined in Eq. (\ref{Lambda}) in
Appendix~\ref{notation}.  Note that $\varepsilon $ is a small positive
constant that prescribes the proper integration contour around the poles
on the real $q_0 $-axis. At the end of calculation, this parameter should
be taken to zero.

By substituting the quark propagator into Eq.~(\ref{Pi-def}),
we arrive at
\ba
\Pi ^{\mu \nu AB}(k) &\simeq & -2 i \pi \alpha N_{f} \delta^{AB}
\int \frac{d q_0 d^2 q}{(2\pi )^{3}} \Bigg[ \frac{\mbox{tr}
\left[ \gamma ^{\mu } \gamma ^{0} \Lambda^{(-)}_{q}
\gamma ^{\nu } \gamma ^{0} \Lambda^{(-)}_{q+k}  \right]}
{ \left( q_0-\epsilon^{-}_{q}
+i \varepsilon \mbox{ sign}(\epsilon^{-}_{q}) \right)
\left( q_0+k_0-\epsilon^{-}_{q+k}
+i \varepsilon \mbox{ sign}(\epsilon^{-}_{q+k}) \right) }
\nonumber \\
&+& \frac{\mbox{tr}
\left[ \gamma ^{\mu } \gamma ^{0} \Lambda^{(+)}_{q}
\gamma ^{\nu } \gamma ^{0} \Lambda^{(-)}_{q+k}  \right]}
{ \left( q_0+\epsilon^{+}_{q} -i \varepsilon \right)
\left( q_0+k_0-\epsilon^{-}_{q+k}
+i \varepsilon \mbox{ sign}(\epsilon^{-}_{q+k}) \right) }
+\frac{\mbox{tr}
\left[ \gamma ^{\mu } \gamma ^{0} \Lambda^{(-)}_{q}
\gamma ^{\nu } \gamma ^{0} \Lambda^{(+)}_{q+k}  \right]}
{ \left( q_0-\epsilon^{-}_{q}
+i \varepsilon \mbox{ sign}(\epsilon^{-}_{q}) \right)
\left( q_0+k_0+\epsilon^{+}_{q+k} -i \varepsilon \right) }
\Bigg].
\label{Pi-long}
\ea
Here we dropped one term in the integrand that contains its both poles in
the upper half of the $q_0$ complex plane. Such a term does not give any
contribution after the integration over $q_0$ is performed.

In the HDL approximation ($|\vec{k}|\ll |\vec{q}|\sim \mu$), the traces
appearing in the last expression are simple (see Appendix~\ref{notation}):
\ba
\mbox{tr}\left[  \gamma ^{\mu } \gamma ^{0} \Lambda^{(-)}_{q}
\gamma ^{\nu } \gamma ^{0} \Lambda^{(-)}_{q+k} \right] &\simeq&
\mbox{tr}\left[  \gamma ^{\mu } \gamma ^{0} \Lambda^{(-)}_{q}
\gamma ^{\nu } \gamma ^{0} \Lambda^{(-)}_{q} \right] =
2\left(g^{\mu 0} + \frac{\vec{q}^{\mu}}{|\vec{q}|}\right)
\left(g^{\nu 0} + \frac{\vec{q}^{\nu}}{|\vec{q}|}\right), \\
\mbox{tr}\left[  \gamma ^{\mu } \gamma ^{0} \Lambda^{(+)}_{q}
\gamma ^{\nu } \gamma ^{0} \Lambda^{(-)}_{q+k} \right] &\simeq&
\mbox{tr}\left[  \gamma ^{\mu } \gamma ^{0} \Lambda^{(+)}_{q}
\gamma ^{\nu } \gamma ^{0} \Lambda^{(-)}_{q} \right] =
-2 g^{\mu \nu } + 2g^{\mu 0} g^{\nu 0}
-2 \frac{\vec{q}^{\mu}}{|\vec{q}|}
\frac{\vec{q}^{\nu}}{|\vec{q}|}, \\
\mbox{tr}\left[  \gamma ^{\mu } \gamma ^{0} \Lambda^{(-)}_{q}
\gamma ^{\nu } \gamma ^{0} \Lambda^{(+)}_{q+k} \right] &\simeq&
\mbox{tr}\left[  \gamma ^{\mu } \gamma ^{0} \Lambda^{(-)}_{q}
\gamma ^{\nu } \gamma ^{0} \Lambda^{(+)}_{q} \right] =
-2 g^{\mu \nu } + 2g^{\mu 0} g^{\nu 0}
-2 \frac{\vec{q}^{\mu}}{|\vec{q}|}
\frac{\vec{q}^{\nu}}{|\vec{q}|}.
\ea

In calculation of $\Pi ^{\mu \nu }(k) $, we first perform the
integration over $q_{0}$ using the residue theorem. The result
reads  (here we drop the overall $\delta^{AB}$)
\ba
\Pi ^{\mu \nu}(k) &\simeq& \frac{\alpha N_{f}}{\pi}
\int d^2 q \left(g^{\mu 0} + \frac{\vec{q}^{\mu}}{|\vec{q}|}\right)
\left(g^{\nu 0} + \frac{\vec{q}^{\nu}}{|\vec{q}|}\right)
\left(
\frac{\theta (\mu -|\vec{q}| ) \theta(|\vec{q}+\vec{k}|-\mu )}
{|\vec{q}|-|\vec{q}+\vec{k}|+k_{0}+ i \varepsilon }
-\frac{\theta (|\vec{q}|-\mu ) \theta(\mu -|\vec{q}+\vec{k}|)}
{ |\vec{q}|-|\vec{q}+\vec{k}|+k_{0}- i \varepsilon }
\right) \nonumber  \\
&+& \frac{\alpha N_{f}}{\pi}
\int d^2 q \left(g^{\mu \nu } - g^{\mu 0} g^{\nu 0}
+ \frac{\vec{q}^{\mu}}{|\vec{q}|} \frac{\vec{q}^{\nu}}{|\vec{q}|}
\right)\left(
\frac{\theta (|\vec{q}|-\mu ) }
{|\vec{q}|+|\vec{q}+\vec{k}|-k_{0}- i \varepsilon }
+\frac{ \theta( |\vec{q}+\vec{k}|-\mu )}
{ |\vec{q}|+|\vec{q}+\vec{k}|+k_{0}- i \varepsilon }
\right)  .
\ea
While the first term in this expression is finite, the second term
contains a linear ultraviolet divergency. As usual, we drop the
ultraviolet divergency from the polarization tensor in the HDL
approximation (it would go away after the the same renormalization
procedure as in the case of zero chemical potential).

The leftover angular integration could be most easily done by going to a
fixed coordinate framework [e.g., such that $\vec {k}=(|\vec{k}|,0)$].
By doing so, we eventually arrive at the following result for the
polarization tensor in the HDL approximation:
\ba
\Pi^{00}(k_0, \vec{k}) & = & \Pi_{l}(k_0, \vec{k}) ,\\
\Pi^{0i}(k_0, \vec{k}) & = &
k_0 \frac{k^i}{|\vec{k}|^2} \Pi_{l} (k_0, \vec{k}) , \\
\Pi^{ij}(k_0, \vec{k}) & = & \left
( \delta^{ij}- \frac{k^i k^j}{|\vec{k}|^2} \right)
\Pi_{t} (k_0,\vec{k})+ \frac{k^i k^j} {|\vec{k}|^2}
\frac{k_0^2}{|\vec{k}|^2} \Pi_{l} (k_0, \vec{k}),
\ea
where
\ba
\Pi_{l}(k_0, \vec{k} )&=& \frac{2 N_{f} \alpha \mu }{\pi}
\int\limits_{0}^{\pi /2} d \varphi
\left( \frac{|\vec{k}| \cos\varphi }
{ k_{0} -|\vec{k}| \cos\varphi + i \varepsilon }
-\frac{|\vec{k} | \cos\varphi }
{ k_{0}+|\vec{k} |\cos\varphi- i \varepsilon }
\right) =M^2\left(
\theta(k^2)\sqrt{\frac{k_0^{2}}{k^2}} -1
-i \theta(-k^2) \sqrt{\frac{k_0^{2}}{-k^2}} \right), 
\label{HDL1}\\
\Pi_{t}(k_0,\vec{k})&=&
\frac{2 N_{f} \alpha \mu }{\pi}
\int\limits_{0}^{\pi /2} d \varphi
\left( \frac{ |\vec{k} |\cos\varphi \sin^{2}\varphi}
{ k_{0} -|\vec{k} | \cos\varphi + i \varepsilon }
-\frac{|\vec{k} | \cos\varphi \sin^{2}\varphi}
{ k_{0}+|\vec{k} |\cos\varphi- i \varepsilon }
+ 2 \cos^{2} \varphi \right)
= M^2-\frac{k^2}{|\vec{k}|^2} \Pi_{l}(k_0,\vec{k})
\label{HDL2}
\ea
with $k^2=k_{0}^{2}-|\vec{k}|^2$.
Here we use the notation $M^2=2 \alpha \mu N_f$ for the Debye mass.  
We should point out that the expression for $\Pi_{l}(k_{0},\vec{k})$,
as written in Eq.~(\ref{HDL1}), is valid only for real values 
of $k_{0}$. The expression for $\Pi_{l}(k_{0},\vec{k})$ as
a function of complex $k_{0}$ has the following form:
\be
\Pi_{l}(k_{0},\vec{k}) = M^{2} \left(
\frac{1}{\sqrt{1-(|\vec{k}|/k_{0})^{2}}} -1\right).
\label{analit-cont}
\ee
Our expressions in Eqs.~(\ref{HDL1}) and (\ref{HDL2}) resemble the results
in the $(3+1)$-dimensional case \cite{Vija,Heinz,Manuel}.  The
corresponding polarization tensor has a nonzero imaginary part for
space-like gluon momenta. This is related to the so-called Landau damping
of the gluon field.

The gluon polarization tensor is transverse
\begin{equation}
k^{\mu} \Pi_{\mu\nu}(k_0, \vec{k})=0,
\end{equation}
and the following representation holds:
\begin{equation}
\Pi_{\mu\nu}=-O^{(1)}_{\mu\nu} \Pi_{t}
+ O^{(2)}_{\mu\nu} \left(\Pi_{t}-M^2\right).
\end{equation}
Here we make use of the projectors of the magnetic, electric and
longitudinal gluon modes introduced in Ref.~\cite{us},
\begin{mathletters}
\ba
O^{(1)}&=& g^{\mu\nu}-u^{\mu} u^{\nu}
+\frac{\vec{k}^{\mu}\vec{k}^{\nu}}{|\vec{k}|^{2}}, \\
O^{(2)}&=& u^{\mu} u^{\nu}
-\frac{\vec{k}^{\mu}\vec{k}^{\nu}}{|\vec{k}|^{2}}
-\frac{k^{\mu}k^{\nu}}{k^{2}}, \\
O^{(3)}&=& \frac{k^{\mu}k^{\nu}}{k^{2}},
\ea
\label{Os}
\end{mathletters}
with $u_{\mu}=(1,0,0,0)$ and
$\vec{k}_{\mu} = k_{\mu} - (u\cdot k) u_{\mu}$.

Then, we rewrite the inverse gluon propagator as
\begin{equation}
\left({\cal D}^{AB}(k_0,\vec{k})\right)_{\mu\nu}^{-1}
=i\delta^{AB} \left(k^2-\Pi_{t}\right) O^{(1)}_{\mu\nu}
+i\delta^{AB} \left(k^2+\Pi_{t}-M^2\right) O^{(2)}_{\mu\nu}
+i\delta^{AB} \frac{k^2}{ \xi } O^{(3)}_{\mu\nu},
\label{inv-D}
\end{equation}
where $\xi$ is a gauge parameter. Then, by making use of the properties of
the projection operators, we invert this expression and arrive at the
final form of the gluon propagator in HDL approximation
\begin{equation}
{\cal D}_{\mu\nu}^{AB}(k_0,\vec{k})=-i\delta^{AB}
\frac{1}{k^2-\Pi_{t}} O^{(1)}_{\mu\nu}
-i\delta^{AB} \frac{1}{k^2+\Pi_{t}-M^2} O^{(2)}_{\mu\nu}
-i\delta^{AB} \frac{ \xi }{k^2} O^{(3)}_{\mu\nu}.
\label{D}
\end{equation}
By using the conventional assumption that the pairing dynamics is
dominated by the momenta $|k_{0}| \ll |\vec{k}|$, we arrive at the
following approximate expression of the propagator in Euclidean
space ($k_{0}=ik_{4}$):
\begin{equation}
i{\cal D}_{\mu\nu}^{AB}(ik_4,\vec{k})\simeq
-\delta^{AB} \frac{|\vec{k}|}{|\vec{k}|^3+ M^2 |k_4| }
O^{(1)}_{\mu\nu}
-\delta^{AB} \frac{1}{k_4^2+|\vec{k}|^2+M^2 }
O^{(2)}_{\mu\nu}
-\delta^{AB} \frac{ \xi }{k_4^2+|\vec{k}|^2} O^{(3)}_{\mu\nu}.
\label{D-long}
\end{equation}

\section{Schwinger-Dyson equation}
\label{SD-sec}

The mechanism of breaking of color symmetry in dense QCD is well known.
The dynamics of quasiparticles around the Fermi surface is affected by the
famous Cooper instability. This causes the perturbative vacuum to
rearrange so that an energy gap is formed in the spectrum of
quasiparticles.  This also is accompanied by the appearance of a color
antitriplet (in the case of three colors) condensate that ``breaks" the
original gauge symmetry $SU(3)_{c}$ down to $SU(2)_{c}$ subgroup. As we
discussed in the preceding section, the corresponding order parameter
leaves the global $SU(2)$ pre-flavor symmetry intact.

In passing we note that because of the Higgs mechanism, five out of the
original eight gluons become massive. The value of the corresponding mass
functions at zero momentum is expected to be of order $\alpha \mu$. This
might suggest that the mentioned five gluons are rather inefficient
in providing interaction in the diquark pairing. If so, one should take
this into account when studying the SD equation. However, we assume that
the Meissner mass function of gluons quickly vanishes (approaching the HDL
approximation) as we go to the region of momenta larger than the
superconducting gap. This is what happens in the $(3+1)$-dimensional model
\cite{us}, and we assume that this should be still valid in $(2+1)$
dimensions.

Our strategy is to use the HDL improved rainbow approximation for
the SD equation and then to check whether it is reliable [as it
happens in the $(3+1)$-dimensional case] for an asymptotically large
chemical potential. The SD equation written in this approximation is:
\begin{eqnarray}
\left(G(p)\right)^{-1}&=&
\left(G^{(0)}(p)\right)^{-1}
+4\pi\alpha \int\frac{d q_{0}d^2 q}{(2\pi)^3}
\sum_{A}\gamma^{\mu}
\left(\begin{array}{cc} T^{A} & 0   \\
       0  & -(T^{A})^{T}
  \end{array}\right)
G(q) \gamma^{\nu}
\left(\begin{array}{cc} T^{A} & 0   \\
       0  & -(T^{A})^{T}
  \end{array}\right)
{\cal D}_{\mu\nu}(q-p),
\label{SD}
\end{eqnarray}
where ${\cal D}_{\mu\nu}(k)$ is the propagator of gluons with the overall
factor $\delta^{AB}$ omitted. We denote the full quark propagator by
$G(p)$, and the perturbative one by $G^{(0)}(p)$.

As is argued in Sec.~\ref{masses}, the order parameter should be a color
anti-triplet and a pre-flavor singlet,
$\varepsilon^{3ab}\langle\bar{\psi}^{C}_{a} T^{2} \psi_{b} \rangle$.
Therefore, the inverse of the full quark propagator should have the
following general structure (neglecting the wave function
renormalization):
\be
\left(G(p)\right)^{-1}=-i\left(\begin{array}{cc}
\pslash+\mu\gamma^{0} & \Delta_{p}\\
\tilde{\Delta}_{p} & \pslash-\mu\gamma^{0} \end{array}\right),
\label{G-inv}
\ee
where $\tilde{\Delta}_{p}=\gamma^0\Delta^{\dagger}_{p}\gamma^0$
and $(\Delta_{p})_{ab} \equiv \varepsilon_{ab3} T^{2} \left(
\Delta^{-}_{p} \Lambda^{-}_{p} + \Delta^{+}_{p} \Lambda^{+}_{p}
\right)$.

By inverting the expression in Eq.~(\ref{G-inv}),
we arrive at the following propagator:
\be
G(p)=i\left(\begin{array}{cc}
R_{1} & \Sigma\\
\tilde{\Sigma} & R_{2} \end{array}\right),
\label{G-one}
\ee
where
\ba
R_{1} &=& {\cal I}_{1} \gamma^{0} \left(
\frac{(p_0+\epsilon^{-}_{p}) \Lambda^{-}_{p} }
{p_0^2-(\epsilon^{-}_{p})^2 -|\Delta^{-}_{p}|^2}
+\frac{(p_0-\epsilon^{+}_{p}) \Lambda^{+}_{p} }
{p_0^2-(\epsilon^{+}_{p})^2 -|\Delta^{+}_{p}|^2}
\right)
+ {\cal I}_{2} \gamma^{0} \left(
\frac{\Lambda^{-}_{p}}{p_0-\epsilon^{-}_{p}}
+\frac{\Lambda^{+}_{p}}{p_0+\epsilon^{+}_{p}}
\right), \\
R_{2} &=& {\cal I}_{1} \gamma^{0} \left(
\frac{(p_0-\epsilon^{-}_{p}) \Lambda^{+}_{p} }
{p_0^2-(\epsilon^{-}_{p})^2 -|\Delta^{-}_{p}|^2}
+\frac{(p_0+\epsilon^{+}_{p}) \Lambda^{-}_{p} }
{p_0^2-(\epsilon^{+}_{p})^2 -|\Delta^{+}_{p}|^2}
\right)
+ {\cal I}_{2} \gamma^{0} \left(
\frac{\Lambda^{+}_{p}}{p_0+\epsilon^{-}_{p}}
+\frac{\Lambda^{-}_{p}}{p_0-\epsilon^{+}_{p}}
\right), \\
\Sigma_{ab} &=& \varepsilon_{ab3} T^{2} \left(
\frac{\Delta^{-}_{p} \Lambda^{+}_{p} }
{p_0^2-(\epsilon^{-}_{p})^2 -|\Delta^{-}_{p}|^2}
+\frac{\Delta^{+}_{p} \Lambda^{-}_{p} }
{p_0^2-(\epsilon^{+}_{p})^2 -|\Delta^{+}_{p}|^2}
\right),
\ea
and $({\cal I}_{1})_{ab} =\delta_{ab}-\delta^{3}_{a}\delta^{3}_{b}$,
$({\cal I}_{2})_{ab}  =\delta^{3}_{a}\delta^{3}_{b}$.

Thus, the gap equation reads
\be
\Delta^{-}_{p} \Lambda^{-}_{p} + \Delta^{+}_{p} \Lambda^{+}_{p}=
-i\frac{8\pi\alpha}{3} \int\frac{d q_{4}d^2 q}{(2\pi)^3}
 \left(
\frac{\gamma^{\mu}\Delta^{-}_{q} \Lambda^{+}_{q} \gamma^{\nu}}
{q_4^2+(\epsilon^{-}_{q})^2 +|\Delta^{-}_{q}|^2}
+\frac{\gamma^{\mu}\Delta^{+}_{q} \Lambda^{-}_{q} \gamma^{\nu}}
{q_4^2+(\epsilon^{+}_{q})^2 +|\Delta^{+}_{q}|^2} \right)
{\cal D}_{\mu\nu}(iq_4-ip_4,\vec{q}-\vec{p}).
\label{SD-1}
\ee
Among two gap functions, $\Delta^{-}_{p} $ and $\Delta^{+}_{p} $,
only the first one plays the role of the gap in the quark
spectrum around the Fermi surface. In addition, the main
contribution in the right hand side of Eq.~(\ref{SD-1})
comes from the first term also proportional to $\Delta^{-}_{p}$.
As a result, the equation for $\Delta^{-}_{p} $ approximately
decouples,
\be
\Delta^{-}_{p}\simeq
-i\frac{4\pi\alpha}{3}
\int\frac{d q_{4}d^2 q}{(2\pi)^3}
 \frac{\Delta^{-}_{q} \mbox{tr} \left[ \gamma^{\mu} \Lambda^{+}_{q}
\gamma^{\nu} \Lambda^{-}_{p} \right] }
{q_4^2+(\epsilon^{-}_{q})^2 +|\Delta^{-}_{q}|^2}
{\cal D}_{\mu\nu}(iq_4-ip_4,\vec{q}-\vec{p}).
\label{SD-2}
\ee
By taking the trace and performing the angular integration (see
Appendix~\ref{notation} for details), we arrive at the following
approximate equation:
\ba
\Delta^{-}_{p} &\simeq &
 \frac{\alpha}{6\pi } \int\frac{d q_{4}d q \Delta^{-}_{q}}
{q_4^2+(\epsilon^{-}_{q})^2 +|\Delta^{-}_{q}|^2}
\Bigg[\frac{1}{\sqrt{ (q-p)^2 +(M^2|q_4-p_4|)^{2/3}}} \nonumber \\
&& + \frac{2}{ \sqrt{(q-p)^2+(q_4-p_4)^2+M^2} }
+\frac{\xi}{ \sqrt{(q-p)^2+(q_4-p_4)^2} }\Bigg].
\label{SD-3}
\ea
To get a analytical estimate for the solution, we assume that the
gap is only a function of the time component of the momentum,
i.e., $\Delta^{-}(p_{4},p) \simeq \Delta^{-}(p_{4},\mu)$. Then,
we can easily perform the integration over $\epsilon^{-}_{q}=
q-\mu$. The approximate result is
\ba
\Delta^{-}(p_{4}) &\simeq &
 \frac{\alpha}{6} \int\frac{d q_{4}\Delta^{-}(q_{4})}
{\sqrt{q_4^2+|\Delta^{-}_{q}|^2}}
\Bigg[\frac{1}{(M^2|q_4-p_4|)^{1/3}}
+ \frac{2}{M}
+\frac{\xi}{|q_4-p_4|}\Bigg].
\label{SD-4}
\ea
Because of the approximations, this last equation makes sense only with
a finite ultraviolet cutoff which should be chosen at the scale of $M$.

Now by making use of the same method as in Ref.~\cite{us} and using the
Landau gauge ($\xi=0$), we straightforwardly obtain an analytical
solution (see Appendix~\ref{AppB} for details):
\be
\Delta^{-}_{0} (p_{4}) = c |\Delta^{-}_{0}|
\left(\frac{|\Delta^{-}_{0}|}{p_{4}}\right)^{1/6}
J_{1}\left(\sqrt{\frac{4\alpha}{(M^{2}p_{4})^{1/3}}}\right),
\quad c \approx 1.6,
\ee
with the overall normalization constant $c$ chosen so that
$\Delta^{-}_{0} (p_{4})|_{p_{4}\approx
|\Delta^{-}_{0}|}=|\Delta^{-}_{0}|$.
The value of the gap itself is equal to
\be
|\Delta^{-}_{0}| \simeq b\frac{\alpha^{3}}{M^{2}}=
b\frac{\alpha^{2}}{2\mu},
\label{result}
\ee
where $b\approx 0.331$. Note that we neglected the effect of electric
gluon modes because their contribution is strongly suppressed compared to
the effect of the magnetic modes.

How reliable is the HDL improved rainbow approximation in
$(2+1)$-dimensional QCD?  Because of a singular nature of the gauge fixing
term in Eq.~(\ref{SD-4}), we see that the above result for the gap is
subject to a large gauge dependent correction in a general covariant
gauge. This indicates that the HDL improved rainbow approximation may not
be reliable. In the next section, we will show that the origin of this
problem in $(2+1)$-dimensional QCD is the phenomenon of nondecoupling of
the dynamics of the fermion pairing from the nonperturbative infrared
dynamics. This phenomenon is intrinsic for lower dimensional field
theories and does not take place in dense QCD in $(3+1)$ dimensions. We
will also discuss a way of solving this problem.

\section{Nondecoupling phenomena in the gap dynamics in 
$(2+1)$-dimensional QCD} 
\label{nondecoupling}

Let us first recall that the HDL improved rainbow approximation is
reliable in $(3+1)$-dimensional dense QCD at asymptotically high quark
densities. This is related to the fact that a large chemical potential
$\mu$ provides a large value of the gap in $(3+1)$ dimensions, namely,
$|\Delta^{-}_{0}| \gg \Lambda_{QCD}$ for a sufficiently large $\mu$
\cite{Son,us,SW2,PR2}. This is enough to justify the use of the one gluon
exchange approximation for the kernel of the SD equation. Indeed, the
integration over the gluon momentum in the SD equation is dynamically cut
at a momentum $k\simeq |\Delta^{-}_{0}| \gg \Lambda_{QCD}$ in infrared,
insuring that the dominant contribution is given by the hard gluons only.
As a result, the diquark pairing dynamics is completely decoupled from the
strong infrared dynamics.

The situation in dense QCD in $(2+1)$ dimensions is different. While the
contribution of hard fermion loops still dominates in the polarization
operator for large $\mu$, the value of the gap (\ref{result}) obtained in
the HDL improved rainbow approximation in the Landau gauge is small.
Therefore, a large region of soft gluons contributes to the SD equation
(moreover, it dominates). It is clear therefore that the one gluon
exchange approximation is not reliable in this case.

What is the physical origin of such a drastic difference between the
$(3+1)$- and $(2+1)$-dimensional cases? We believe it is a diminishing
role of the Fermi surface in the fermion pairing dynamics in lower
dimensions. Indeed, the main effect of the Fermi surface in the fermion
pairing is reducing the initial dimension $d+1$ to the effective dimension
$1+1$. This is crucial for $d=3$. However, for example, for $d=1$ there is
no such a reduction at all: there is $1+1$ dimensional dynamics from the
outset!\footnote{In passing, we note that in the $1+1$ case, in the HDL
approximation, the Debye screening mass is proportional to $\alpha_{2}$
(where $\alpha_{2}$ is the gauge coupling) and is independent of $\mu$ at
all!  The Debye mass actually coincides with the famous Schwinger mass,
appearing for $\mu$=0 in $1+1$ [see also a discussion in
Ref.~\cite{PRAZ}]. This observation certainly implies the breakdown of the
HDL expansion in $1+1$ dimensional QCD.} The present case with $d=2$ is
intermediate, and our analysis shows how different it is from the $d=3$
case.

The importance of the region of soft gluon momenta in the SD equation
implies that the pairing dynamics in $(2+1)$-dimensional QCD is intimately
connected with the strong infrared dynamics. In fact, there is no
decoupling of these two dynamics even for asymptotically large values of
the chemical potential.

Let us show that this point is directly related to the existence of an
upper bound for the value of the gap in the $(2+1)$-dimensional QCD. In
particular, $|\Delta^{-}_{0}|$ cannot be much larger than $\alpha$.
Indeed, if the value of the gap were much larger than $\alpha$, then the
HDL improved rainbow approximation would be reliable. But the only
solution of the corresponding equation is given in Eq.~(\ref{result}),
showing that $|\Delta^{-}_{0}| \ll \alpha$. Thus, the assumption of large
$|\Delta^{-}_{0}|$ is self-contradictory.

Therefore in order to solve the problem for large $\mu$ in $(2+1)$
dimension, one should put the non-perturbative infrared dynamics under
control. Does it imply that expression (\ref{result}) obtained in the
HDL improved rainbow approximation is irrelevant? We do not think so: we
will argue that it can be qualitatively correct in the case of a large
number of fermion flavors $N_{f}$.

The fact that the $1/N_{f}$ expansion can be useful in $(2+1)$-dimensional
QED and QCD was recognized long ago \cite{P84,ABKW,ANW,AN}.  The point is
that because these models are renormalizable theories, they are
asymptotically free for any number $N_{f}$. Though being very
nontrivial, the $1/N_{f}$ expansion is helpful in putting under control
the infrared dynamics. The crucial point is the selection of a ``right" 
gauge in the leading order in $1/N_{f}$. In particular, appropriate Ward
identities have to be satisfied in that gauge. In other gauges, the result
can be found by gauge-transforming Green functions from the ``right"
gauge to those gauges. The description of this very nontrivial procedure
can be found in Ref.~\cite{KN}. For our purposes, it is enough to know  
that though the ``right" gauge is different from the Landau one, the
results obtained in the improved rainbow approximation in the latter are
qualitatively reliable: its special role amongst other covariant gauges is
connected with the approximate validity of the Ward identities in the  
improved rainbow approximation using the bare vertices.

It is reasonable to assume that the Landau gauge is special in our problem
either. Indeed, let us recall that, because of the Miessner effect, the  
quark gluon vertices should necessarily receive non-perturbative pole   
corrections \cite{bs-all}.  This is a rather general and a model
independent property that is related to the Ward identities \cite{JJ,CN} 
for the vortex function.  Following Ref.~\cite{bs-all}, one can show that
in a general covariant gauge the contribution of those pole terms in the  
SD equation is large, thus destroying the validity of the approximation
with bare vertices.  The solution to a similar problem was suggested long
time ago by Cornwall and Norton \cite{CN}. The idea is to choose the gauge
where the approximation of the SD equation with bare vertices is
self-consistent. In their model, it was the Landau gauge. With this gauge
choice, it was possible to show that potentially dangerous contributions
cancel from the SD equation. This is possible due to the fact that the
pole contribution to the vertex is pure longitudinal. The Landau gauge is
special because the gluon propagator is completely transverse. Thus, when
used in the SD equation, such gluon propagator annihilates the dangerous  
infrared poles.

Using the same arguments, we see that the Landau gauge is the best gauge  
among the general covariant gauges in the problem at hand too. We should
note, however, that our situation differs from that in Ref.~\cite{CN}. The
Lorentz symmetry is broken in dense QCD by a nonzero chemical potential.
As a result, the pole structure of the quark gluon vertex is proportional
to $\tilde{p}^{\mu}-\tilde{q}^{\mu} = (p^{0}-q^{0},(\vec{p}-\vec{q})/3)$,
rather than $p^{\mu}-q^{\mu}$ as in Ref.~\cite{CN}. While the four-vector
$\tilde{p}^{\mu}-\tilde{q}^{\mu}$ is indeed annihilated by the magnetic   
modes of gluons, it is only partially annihilated by the electric ones. It
is fortunate that the electric gluon modes are screened much stronger and
their effect on diquark pairing is negligible in the model at hand. So, we
could still justify the use of the Landau gauge, in which all most
dangerous infrared contributions are canceled from the SD equation.

These arguments lead us to believe that expression (\ref{result}) can be
qualitatively reliable for large $N_{f}$. 
We rewrite the result in the following form:
\be
|\Delta^{-}_{0}| \simeq
\bar{b}\frac{\bar{\alpha}^{2}}{2N_{f}^{2}\mu},
\label{Nresult}
\ee
where $\bar{\alpha}=N_{f}\alpha$ and the constant $\bar{b}$ is of order
one. We got this estimate by making use of conventional large $N_f$
rescaling of the gauge coupling. Though a color-flavor structure of the
condensate can vary with $N_{f}$, we expect that the magnitude of the gap
should be similar for different, but same order, large $N_{f}$.

\section{Conclusion}
\label{conclusion}

We studied the dynamical generation of the color superconducting order
parameter in $(2+1)$-dimensional QCD at asymptotically large baryon
density and zero temperature. For this purpose, we used the conventional
method of the Schwinger-Dyson equation in the HDL improved rainbow
approximation. It was found that the order parameter in this theory is
similar to that in the $S2C$ phase of the $(3+1)$-dimensional two flavor
QCD.

By solving the SD equation analytically, we obtained an approximate
solution for the color superconducting gap. The parametric dependence of
the result on the coupling constant is given by a power law. This is
qualitatively different from the situation in the $(3+1)$-dimensional QCD,
where the expression for the gap is exponentially small for small values
of the coupling constant \cite{Son,us,SW2,PR2}.

The analysis of the solution of the SD equation revealed a qualitatively
new [with respect to the $(3+1)$-dimensional case] phenomenon:
nondecoupling of the dynamics of the fermion pairing from the
nonperturbative infrared one, even in the case of asymptotically large
values of the chemical potential. We believe that this phenomenon is
common in lower dimensional models and deserves further study.

It would be also interesting to study the interplay between 
the dynamics of color and chiral condensates in $(2+1)$-dimensional 
QCD. It is known that, at zero quark density, a chiral 
condensate can occur only if the number of quark flavors is less 
than a critical value $N_{f}^{cr}$, which is a function of the 
number of colors $N_{c}$ \cite{ANW,AN}. An interesting question 
is whether, in this case, there is a phase transition which 
restores chiral symmetry at a critical value of the chemical
potential and how the character of this phase transition depends 
on the values of $N_{f}$ and $N_{c}$.

In the analysis of the SD equation we assumed that the Meissner effect is
mostly irrelevant for the pairing dynamics in dense quark matter.  While
this was well justified in $(3+1)$-dimensional QCD at asymptotic densities
\cite{us,SW2,Ris01}, it remains an open issue in the $(2+1)$-dimensional
case. Indeed, our present analysis suggests that the dominant contribution
to the gap equation comes from the region of momenta not much larger than
$|\Delta^{-}|$. The Meissner effect, therefore, could play a significant
role in modifying the gluon interaction in the diquark channel. It would
be interesting to study this problem in more detail.

It is interesting to point out that, because of close similarity of the
model studied here with that of the $(3+1)$-dimensional two flavor QCD,
one should also expect the appearance of five light ($M_{dq}\ll
|\Delta_{0}|$) pseudo-Nambu-Goldstone diquarks in the low energy spectrum
\cite{bs-all}. The argument for the existence of such states is
essentially the same as in the $(3+1)$-dimensional case, see
Ref.~\cite{bs-all}. The only difference should appear in the hierarchy of
the scales.

At last, we would like to mention that some relativistic
$(2+1)$-dimensional gauge models are used by some theorists for describing
high temperature superconductivity in cuprates \cite{Franz}. It would be
interesting to see whether the color superconductivity could anyhow be
related to the ordinary superconductivity in planar systems.

\begin{acknowledgments} 
V.A.M., G.S. and L.C.R.W. thank Professor H.~Minakata for his hospitality
at the T.M.U. during the Yale T.M.U. workshop on gauge theories. They also
thank T.~Appelquist for valuable discussions. The work of G.W.S. was
supported by the Natural Sciences and Engineering Research Council of
Canada. I.A.S. would like to thank V.~Gusynin for useful discussions and
comments. The work of I.A.S. was supported by the U.S. Department of
Energy Grant No.~DE-FG02-87ER40328 and partially by Grant-in-Aid of Japan
Society for the Promotion of Science (JSPS) No.~11695030. The work of
L.C.R.W. was supported by the U.S. Department of Energy Grant No.
DE-FG02-84ER40153.  
\end{acknowledgments}

\section*{Note}
While writing our paper, we learned that a partially overlapping
study is being done by Prashanth Jaikumar and Ismail Zahed.

\appendix

\section{Notation and useful formulas}
\label{notation}

We use the four-dimensional representation of the Dirac $\gamma$-matrices.
In addition to the three matrices $\gamma^{\mu}$ ($\mu=0,1,2$) there are
two more matrices that anticommute with $\gamma^{\mu}$ as well as between
themselves, $\gamma^{3}$ and $\gamma^{5}$ [here we use the same notation
for the $\gamma$-matrices as one would use in the $(3+1)$-dimensional
case].

While studying the color superconductivity in the model, it is
convenient to introduce the charge conjugate spinors as follows:
\ba
\psi^{C}(x)=C\bar{\psi}^{T}(x), &\qquad &
\bar{\psi}(x)=-\left[\psi^{C}(x)\right]^{T} C^\dagger ,\\
\bar{\psi}^{C}(x)=-\psi^{T}(x) C^\dagger ,  &\qquad&
\psi(x)=C\left[\bar{\psi}^{C}(x)\right]^{T},
\ea
where $C$ is a unitary charge conjugation matrix, defined
by $C^{-1} \gamma_{\mu}C =-\gamma_{\mu}^{T}$ and $C=-C^{T}$.

Note that in momentum space,
\ba
\psi^{C}(p)=C\bar{\psi}^{T}(-p) , &\qquad &
\bar{\psi}(p)=-\left[\psi^{C}(-p)\right]^{T} C^\dagger , \\
\bar{\psi}^{C}(p)=-\psi^{T}(-p) C^\dagger  , &\qquad&
\psi(p)=C\left[\bar{\psi}^{C}(-p)\right]^{T}.
\ea
The Lagrangian density of QCD could be written as
\ba
{\cal L}_{QCD}&=& \frac{1}{2} \bar{\psi}
\left(\pslash +\mu \gamma^0 \right) \psi
+ \frac{1}{2} \bar{\psi}^{Ci}
\left(\pslash -\mu \gamma^0 \right) \psi^{C}
+ \frac{1}{2} \bar{\psi} \gamma^{\mu }
\hat{A}_{\mu } \psi^{i}
- \frac{1}{2} \bar{\psi}^{C} \gamma^{\mu }
\hat{A}^{T}_{\mu } \psi^{C}
-\frac{1}{16\pi\alpha} \mbox{Tr} \left(\hat{F}_{\mu \nu }
\hat{F}^{\mu \nu } \right). 
\label{L-QCD}
\ea
The eight component Majorana spinor is defined by
\be
\Psi=\frac{1}{\sqrt{2}} \left(
\begin{array}{c} \psi \\ \psi^{C} \end{array} \right) .
\ee
In this new notation, the bare fermion propagator reads
\begin{equation}
\left(G^{(0)}(p)\right)^{-1}=-i\left(\begin{array}{cc}
\pslash +\mu \gamma^{0} & 0\\ 0 & \pslash -\mu \gamma^{0}
\end{array}\right) .
\label{bare-G}
\end{equation}
Note the following convenient representations for the
diagonal elements of this propagator:
\ba
\pslash+\mu\gamma^{0} &=& \gamma^{0}\left[
(p_0-\epsilon^{-}_{p}) \Lambda^{(+)}_{p}
+(p_0+\epsilon^{+}_{p}) \Lambda^{(-)}_{p} \right], \label{a-1}\\
\pslash-\mu\gamma^{0} &=& \gamma^{0}\left[
(p_0-\epsilon^{+}_{p}) \Lambda^{(+)}_{p}
+(p_0+\epsilon^{-}_{p}) \Lambda^{(-)}_{p} \right], \label{a-2}\\
\left(\pslash+\mu\gamma^{0}\right)^{-1} &=& \gamma^{0}\left[
\frac{1}{p_0+\epsilon^{+}_{p}} \Lambda^{(+)}_{p}
+\frac{1}{p_0-\epsilon^{-}_{p}} \Lambda^{(-)}_{p} \right],
\label{a-3}\\
\left(\pslash-\mu\gamma^{0}\right)^{-1} &=& \gamma^{0}\left[
\frac{1}{p_0+\epsilon^{-}_{p}} \Lambda^{(+)}_{p}
+\frac{1}{p_0-\epsilon^{+}_{p}} \Lambda^{(-)}_{p} \right],
\label{a-4}
\ea
where $\epsilon_{p} ^{\pm }=|\vec{p}|\pm \mu $ and
the ``on-shell" projectors are
\ba
\Lambda^{(\pm )}_{p}=\frac{1}{2}
\left( 1 \pm \frac{\vec{\alpha}\cdot \vec{p}}{|\vec{p}|} \right),
\quad \vec{\alpha} \equiv \gamma^{0} \vec{\gamma}.
\label{Lambda}
\ea
The bare quark-quark-gluon vertex, in the eight component notation,
is also a matrix,
\be
\gamma^{\mu}\left(\begin{array}{cc}
T^{A} & 0 \\
0  &  -(T^{A})^{T}
\end{array}\right).
\ee
While working with the SD equation in Sec.~\ref{SD-sec}, we encounter a
few Dirac traces. Here are the results for most general ones:
\ba
\mbox{~tr}\left[\gamma^{\mu} \Lambda^{(e)}_{p}
\gamma^{\nu} \Lambda^{(e')}_{q}\right] &=&
g^{\mu\nu} (1 + e e^{\prime} t )
-2 e e^{\prime} g^{\mu 0}g^{\nu 0} t + e e^{\prime}
\frac{\vec{q}^{\mu}\vec{p}^{\nu}+\vec{q}^{\nu}\vec{p}^{\mu}}
{|\vec{q}| |\vec{p}|}
+ \ldots,
\label{tr1}\\
\mbox{~tr}\left[\gamma^{\mu} \gamma^0 \Lambda^{(e)}_{p}
\gamma^{\nu} \gamma^0 \Lambda^{(e^{\prime})}_{q}\right]
&=& -g^{\mu\nu} (1-e e' t )
+\left(g^{\mu 0} - e^{\prime} \frac{\vec{q}^{\mu}}{|\vec{q}|}\right)
\left(g^{\nu 0} - e \frac{\vec{p}^{\nu}}{|\vec{p}|}\right)
\nonumber \\
&&+\left(g^{\mu 0} - e \frac{\vec{p}^{\mu}}{|\vec{p}|}\right)
\left(g^{\nu 0} - e^{\prime}\frac{\vec{q}^{\nu}}{|\vec{q}|}\right)
+ \ldots,
\label{tr2}
\ea
where $e,e^{\prime}=\pm 1$, $t =\cos\varphi $ is the cosine of the angle
between two-vectors $\vec{q}$ and $\vec{p}$, and irrelevant antisymmetric
terms are denoted by ellipsis.

By making use of the traces above, we calculate the following
expressions which appear in the SD equation:
\begin{mathletters}
\ba
O^{(1)}_{\mu\nu} \mbox{~tr}\left[\gamma^{\mu} \Lambda^{(+)}_{q}
\gamma^{\nu}\Lambda^{(-)}_{p}\right] &=& (1-t ) \frac{(q+p)^2 }
{q^2+p^2-2q pt }, \label{O11}\\
O^{(2)}_{\mu\nu} \mbox{~tr}\left[\gamma^{\mu}\Lambda^{(+)}_{q}
\gamma^{\nu}\Lambda^{(-)}_{p}\right] &=&
(1+t ) \frac{(q-p)^2}{q^2+p^2-2q pt }
+(1-t^2) \frac{2q p }{q^2+p^2-2q pt+(q_4-p_4)^2}, \label{O12}\\
O^{(3)}_{\mu\nu} \mbox{~tr}\left[\gamma^{\mu}\Lambda^{(+)}_{q}
\gamma^{\nu}\Lambda^{(-)}_{p}\right] &=& (1+t )
\frac{(q-p)^2+(q_4-p_4)^2}
{q^2+p^2-2q pt +(q_4-p_4)^2}, \label{O13}
\ea
\end{mathletters}
where $q\equiv |\vec{q}|$, $p\equiv |\vec{p}|$, $q_4\equiv -i q_0$
and $p_4\equiv -i p_0$. Making use of these expression, we perform
the angular integration that appears in the SD equation,
\ba
q \int d\varphi
{\cal D}_{\mu\nu}(q-p) \mbox{~tr}\left[ \gamma^{\mu}
\Lambda^{(-)}_{p} \gamma^{\nu} \Lambda^{(+)}_{q}
\right] &\approx & i \pi \Bigg[
\frac{1}{\sqrt{ (q-p)^2+(M^2\omega)^{2/3} } }
+ \frac{2}{ \sqrt{(q-p)^2+\omega^2+M^2} } \nonumber\\
&&+\frac{\xi}{ \sqrt{(q-p)^2+\omega^2} }\Bigg].
\label{angle1}
\ea
Here $M^2=2\alpha\mu N_{f}$ and $\omega=|q_4-p_4|$. Although we
performed the angular integration exactly, we dropped all subleading
terms on the right hand of Eq.~(\ref{angle1}), assuming that both $q$ and
$p$ remain in the vicinity of the Fermi surface.

\section{Approximate solution of the gap equation}
\label{AppB}

In this Appendix we present an approximate analytical solution to
the gap equation (\ref{SD-4}). In order to reduce the integral gap
equation to a differential one, we introduce an infrared cutoff in the
integral at the scale of $|\Delta_{0}^{-}|$, and we approximate the
kernel in the integrand by its asymptotes:
\be
\frac{1}{\sqrt{q_4^2+|\Delta^{-}_{q}|^2}}
\frac{1}{(M^2|q_4-p_4|)^{1/3}} = \left\{
    \begin{array}{lll}
    \frac{1}{(M^2|p_4|)^{1/3} q_4}, &
    \quad \mbox{for} \quad &  q_4 \leq p_4 , \\
    \frac{1}{M^{2/3} |q_4|^{4/3}}, &
    \quad \mbox{for} \quad & q_4 \geq p_4 .
\end{array} \right.
\label{B-1}
\ee
We then arrive at the following equation:
\be
\Delta^{-}(p_{4}) \simeq
\frac{\nu^{1/3}}{3|p_4|^{1/3}}
\int_{|\Delta_{0}^{-}|}^{p_4}
\frac{d q_{4}}{q_4} \Delta^{-}(q_{4})
+ \frac{\nu^{1/3}}{3} \int_{p_4}^{\infty}
\frac{d q_{4}}{|q_4|^{4/3}}\Delta^{-}(q_{4}),
\label{B-2}
\ee
where
\be
\nu \equiv \frac{\alpha^{3}}{M^{2}}.
\ee
It is straightforward to check that the integral equation is equivalent to
the following second order differential equation
\be
\frac{d^{2} \Delta^{-}(p_{4})}{dp_{4}^{2}}
+\frac{4}{3 p_{4}}
\frac{d \Delta^{-}(p_{4})}{dp_{4}}
+\frac{\nu^{1/3}}{9 p_{4}^{7/3}} \Delta^{-}(p_{4}) =0,
\label{B-3}
\ee
along with the infrared and ultraviolet boundary conditions:
\ba
\left .
\frac{d \Delta^{-}(p_{4})}{dp_{4}}
\right|_{p_{4}=|\Delta^{-}_{0}|} = 0 ,
\label{B-IR}\\
\left .\left(
3 p_{4} \frac{d \Delta^{-}(p_{4})}{dp_{4}}
+\Delta^{-}(p_{4}) \right) \right|_{p_{4}=\infty} = 0 .
\label{B-UV}
\ea
The general solution to the differential equation (\ref{B-3})
is given in terms of Bessel functions,
\be
\Delta^{-} (p_{4}) = |\Delta^{-}_{0}| \left[ C_{1}
\left(\frac{|\Delta^{-}_{0}|}{p_{4}}\right)^{1/6}
J_{1}\left(2\frac{\nu^{1/6}}{p_{4}^{1/6}}\right)
+C_{2} \left(\frac{|\Delta^{-}_{0}|}{p_{4}}\right)^{1/6}
Y_{1}\left(2\frac{\nu^{1/6}}{p_{4}^{1/6}}\right)
\right]. \label{B-4}
\ee
In order to satisfy the ultraviolet boundary condition in
Eq.~(\ref{B-UV}), we must choose $C_{2}=0$. In passing we note, however,
that when a finite ultraviolet cutoff is used in the gap equation
(\ref{B-2}) (one might choose, for example, $q_{4}=M$), the integration
constant $C_{2}$ will be nonzero,
\be
C_{2}\simeq \frac{\pi}{2} \left(\frac{\nu}{M}\right)^{2/3} C_{1}.
\ee
As is easy to check, the effect of this nonzero constant is negligible
in the leading order approximation.

Now, by satisfying the infrared boundary condition, we arrive at the
following equation:
\be
J_{1}\left(2\frac{\nu^{1/6}}{|\Delta^{-}_{0}|^{1/6}}\right)
=\left(\frac{\nu}{|\Delta^{-}_{0}|}\right)^{1/6}
J_{2}\left(2\frac{\nu^{1/6}}{|\Delta^{-}_{0}|^{1/6}}\right)
-\left(\frac{\nu}{|\Delta^{-}_{0}|}\right)^{1/6}
J_{0}\left(2\frac{\nu^{1/6}}{|\Delta^{-}_{0}|^{1/6}}\right).
\ee
This equation has infinitely many solutions, which correspond to different
solutions to the gap equation. We must choose a single solution which gives
the gap function without nodes. As one could check, this is also equivalent
to choosing the solution with the largest value of the gap. Thus, we obtain
$|\Delta^{-}_{0}| = b \nu$ where $b \approx 0.331$.

\end{document}